\documentclass[aps,prd,twocolumn,groupedaddress]{revtex4-2}
\usepackage{graphicx}
\usepackage{xcolor}
\usepackage{hyperref}
\begin{document}
\title{0.5 eV QCD Axion Cosmology}
\author{Noah Bray-Ali}
\affiliation{Science Synergy, Los Angeles, CA 90045}
\email{nbrayali@gmail.com}
\date{November 13, 2026}





\begin{abstract}
The best available determination of the present expansion rate of the universe using late-universe observations, by the SH0ES collaboration in 2025, differs by more than seven standard deviations from the value of the Hubble constant determined by the Planck collaboration in 2018 using early-universe observations and the standard cold dark matter cosmology with cosmological constant, a discrepancy known as the Hubble tension. Within a spatially flat, isotropic, and homogeneous expanding-universe solution of the field equations of general relativity with cosmological constant, the SH0ES value for the Hubble constant implies roughly twice as many baryons as the standard cosmology, provided that one retains the Planck values for the energy density of cold dark matter in the present universe and for the cosmological constant. A novel cosmology is proposed --- in terms of cooling dark matter made of quantum chromodynamic (QCD) axions with present number density six times that of the photons in the cosmic microwave background --- which realizes this straightforward, doubled-baryons scenario for resolving the Hubble tension.
\end{abstract}
\maketitle


		
	
	
	\section{Introduction}
	\label{sec:intro}
	
	In 2022 the tension between late-universe observations \citep{riess2021} of the present expansion rate, known as the Hubble constant, $H_0$, and the prediction for $H_0$ within the standard cold dark matter cosmology with cosmological constant (${\rm \Lambda CDM}$), based on early-universe observations \citep{planck2020,act2020}, reached the conventional $5.0\sigma$ threshold for the discovery of a new phenomenon in astrophysics. Several years before this conventional threshold was reached, astrophysicists were already stimulated by the accumulating evidence for the existence of a significant Hubble tension, and they began suggesting innovative alternative cosmologies that might reconcile early- and late-universe observations \citep{poulin2019}. In 2024 the Hubble tension was confirmed on the late-universe side with James Webb Space Telescope (JWST) observations that rejected unrecognized crowding of Cepheid variable star photometry as an explanation of the Hubble tension at $8\sigma$ confidence \citep{riess2024_crowding}, and that validated the Hubble Space Telescope (HST) distance measurements on which the tension rests \citep{riess2024_selection}.
	
	On the early-universe side, the Dark Energy Spectroscopic Instrument (DESI) first-year observations of the peak in the galaxy-galaxy two-point correlation function confirmed the Hubble tension in 2025, provided the results are interpreted within the standard $\Lambda{\rm CDM}$ cosmology \citep{desi2025_hubble}. Similarly, in 2023 an independent combination of ground-based and space-based observations of the cosmic microwave background (CMB) power spectrum of angular fluctuations confirmed the tension, again within the framework of $\Lambda{\rm CDM}$ \citep{spt2023}. In 2024 a comprehensive book-length review concluded that the Hubble tension remains an unsolved mystery \citep{riess2024_book}.
	
	Here we propose a novel cosmology --- in terms of cooling dark matter made of quantum chromodynamic (QCD) axions \cite{wilczek1978,weinberg1978} with rest-mass energy around 0.5 eV and with present number density six times the number density of photons in the cosmic microwave background (CMB) --- in which, as in $\Lambda{\rm CDM}$, space grows with time in a manner described by a spatially flat, homogeneous, isotropic, expanding-universe solution to the field equations of general relativity with cosmological constant term \citep[pg.~62, 65, 70--76, 265--268]{peebles1993}. Remarkably, this 0.5 eV QCD axion cosmology yields a high-precision prediction from first-principles for $H_0$ (See Section \ref{sec:hubble}):
	\begin{eqnarray}
	\label{eq:hubble}
	H_0 = 72.067~3~(79)~{\rm km\cdot s^{-1}\cdot Mpc^{-1}}.
\end{eqnarray}     
	Intriguingly, this high-precision prediction from first-principles for the value of $H_0$ within the proposed cosmology {\em does not} differ sharply from the best present value from late-universe observations, $H_0({\rm SH0ES})$, determined by the SH0ES collaboration in 2025 \citep{riess2025}, using Type Ia supernovae calibrated both by Cepheid variables and by the tip of the red giant branch (TRGB), and relying on both JWST and HST data: 
	\begin{eqnarray}
		\label{eq:shoes}
		H_0({\rm SH0ES}) = 73.18~(88)~{\rm km\cdot s^{-1}\cdot Mpc^{-1}}.
	\end{eqnarray} 
	
	Perhaps even more remarkably, however, the proposed cosmology also yields predictions --- again from first-principles and with high-precision --- for the values of the cosmological parameters of the cosmological constant, $\Omega_{\Lambda}$, and the present energy density of cold dark matter, $\Omega_Ah^2$ (See section \ref{sec:hubble}):
	\begin{eqnarray}
		\label{eq:axion_lambda}
		\Omega_Ah^2=0.117~952~(78),~~~~~\Omega_{\Lambda}=0.678~08~(15).
	\end{eqnarray}
	Strikingly, these predicted values do not sharply disagree with the best present {\em early-universe} observational values, $\Omega_{\Lambda}(Pl)$ and $\Omega_Ah^2(Pl)$, determined by the {\em Planck} collaboration in 2018 from analysis of cosmic microwave background fluctuations, baryon-acoustic oscillations, and cosmic shear \citep{planck_baseline,planck2020}:
	\begin{eqnarray}
		\label{eq:axion_lambda_planck}
				\Omega_Ah^2(Pl)=0.118~82~(86),~\Omega_{\Lambda}(Pl)=0.692~(5).
	\end{eqnarray}
	Instead, the sharp difference --- with more than $10\sigma$ significance --- between the predicted value for $H_0$, given in Eq.~(\ref{eq:hubble}), and the {\em Planck} value, $H_0(Pl)=67.9~(4)$ \citep{planck_baseline,planck2020}, is due to the roughly doubled size for the predicted baryon abundance, $\Omega_Bh^2$, compared to the {\em Planck} value, $\Omega_Bh^2(Pl)$, which is compensated in the fits to early-universe observations by a sharply larger value ---with nearly $9\sigma$ significance --- for the predicted spectral index of primordial energy density fluctuations, $n_s$, compared to the {\em Planck} value, $n_s(Pl)$ \citep{planck_baseline,planck2020} (See section \ref{sec:hubble}):
	\begin{eqnarray}
		\label{eq:ns-baryon}
		\Omega_Bh^2&=&0.049~243~(32),~~~~~~n_s=1\nonumber\\
	    \Omega_Bh^2(Pl)&=&0.022~46~(13),~~n_s(Pl)=0.967~7~(37).
	\end{eqnarray} 
	
	In this way, a straightforward resolution of the Hubble Tension is proposed --- within 0.5 eV QCD axion cosmology --- that embraces both the late-universe value for the Hubble constant, given in Eq.~(\ref{eq:shoes}), and the early-universe values for the cosmological constant and cold dark matter abundance, given in Eq.~(\ref{eq:axion_lambda_planck}). The key point motivating this doubled-baryon scenario is that the values for the baryon abundance and the spectral index are degenerate with one another in the fits to early-universe observations \citep{planck2020}. By simultaneously raising both values, the quality of the fit to early-universe observations (CMB fluctuations, BAO, and cosmic shear) stays roughly the same as in the standard ${\rm \Lambda CDM}$ cosmology.
	
	The paper is organized as follows. Section \ref{sec:cosmax} addresses the timing of the onset of large-scale structure formation within the proposed cosmology and shows that it comes out about the same as in ${\rm \Lambda CDM}$, despite the very different physical pictures for the nature of the onset in the two cosmologies. Section \ref{sec:bbn} addresses the timing of the onset of Big Bang nucleosynthesis (BBN) within the proposed 0.5 eV QCD axion cosmology and shows that this timing combines with the doubled baryons to yield roughly the same primordial deuterium abundance as $\Lambda{\rm CDM}$ --- in agreement with the best present observational determination \cite{cooke2018} --- and that this combination yields roughly ten per cent more helium than the standard BBN predictions \cite{parthenope2021,primat2018}. Section \ref{sec:bkg} presents the physical picture for the nature of the 0.5 eV QCD axions that form the dark matter within the proposed cosmology. Section \ref{sec:hubble} derives the values of the cosmological parameters. Finally, section \ref{sec:conc} summarizes the paper, presents an outlook for future work, and discusses astrophysical constraints related to observations of globular clusters, red giants, white dwarfs, pulsars, the neutrino pulse from supernova 1987A, and the Milky Way interstellar mid-infrared background.

	\section{Large-Scale Structure Formation}
	\label{sec:cosmax}
	\begin{figure}
		\centering
		\includegraphics[width=1.0\linewidth]{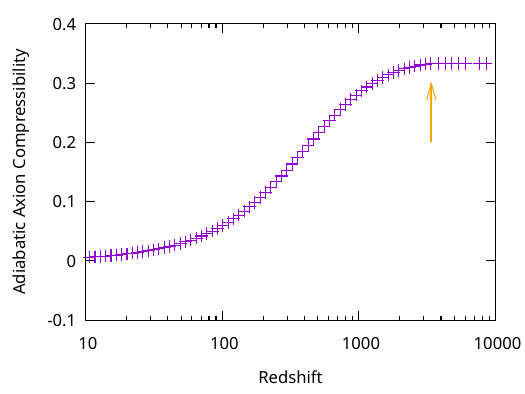}
		\label{fig:cosmax}
		\caption{Cooling of 0.5 eV QCD axion dark matter in the early universe shows a cross-over in the adiabatic compressibility, $dp_A/du_A$ (violet pluses, \textcolor{violet}{+}), from the radiation-like value, $dp_A/du_A\approx1/3$, at large redshift, to the matter-like value, $dp_A/du_A\approx0$, at small redshift, starting right around the redshift of ``matter-radiation equality,'' $z_{\rm eq}=3376~(19)$ (vertical orange arrow), predicted by the standard cold dark matter cosmology with cosmological constant ($\Lambda{\rm CDM}$) using observations of the early universe \citep{planck_baseline,planck2020}.}
	\end{figure}
	
	The formation of large-scale astrophysical structures in the early universe is a key test of any proposal for the nature of dark matter, dark energy, and the Big Bang, including 0.5 eV QCD axion cosmology. Historically, the now-standard cold dark matter cosmology first emerged in the context of concerns about the relative {\em lack} of large-scale anisotropy in the cosmic microwave background (CMB), compared to expectations based on the then-standard baryon-based Big Bang cosmology \cite[pg.~301--302]{peebles2020}. In fact, Peebles first proposed a hot dark matter cosmology \citep{peebles1982hdm}, before turning to cold dark matter \citep{peebles1982cdm}, with the shared goal, in both cases, of ``washing out'' large-scale CMB anisotropy and, in this way, resolving the tension between the then-standard cosmology and early-universe observations \cite{fixsen1983}.
	
	Figure 1 shows the cross-over in the equation of state that relates the pressure, $p_A(z)$, and the energy density, $u_A(z)$, for 0.5 eV QCD axion dark matter in thermal equilibrium with the cosmic background radiation at temperature, $(1+z)T_{\gamma}$ \citep[pgs.~134--135]{peebles1993}, where $z$ is the redshift and $T_{\gamma}=2.7255~(6)~{\rm K}$ \cite{fixsen} is the temperature of the CMB. At large redshift, $z\gg10^4$, the adiabatic axion compressibility, $dp_A/du_A\approx1/3$, takes the radiation-like value, reflecting a light-like equation of state, $p_A(z)=u_A(z)/3$ \citep[pg.~139]{peebles1993}, while at small redshift, $z\ll 10^2$, the value crosses over to matter-like, $dp_A/du_A\approx0$. Crucially, the cross-over begins at roughly the best present estimate of the redshift for ``matter-radiation equality,'' $z_{\rm eq}=3376~(19)$ \citep{planck_baseline,planck2020},  within the standard cold dark matter cosmology.
	
	As the 0.5 eV QCD axion dark matter equation of state crosses over to matter-like at small redshifts from radiation-like at large redshifts, starting around $z_{\rm eq}$, as shown in Fig.~2, the dynamics of the expansion change to ``matter-dominated'' from ``radiation-dominated.'' Indeed, with six axions for each photon made in the cosmic background radiation (See Section \ref{sec:hubble} and Appendix \ref{sec:appendix_A}), axions dominate the expansion dynamics on both sides of ``matter-radiation equality.'' The cross-over in the axion equation of state (Fig.~2), beginning around $z_{\rm eq}$, drives the cross-over in the character of the expansion which then drives the large-scale structure formation seen by early-universe observations, just as the change from ``radiation-domination'' to ``matter-domination'' drives such structure to form in the standard cold dark matter cosmology \citep[pgs.~624--625]{peebles1993}.
	
	The rough agreement, shown in Fig.~2, between the best present observational value of the redshift where large-scale structure formation begins, $z_{\rm eq}$, and the predicted value for this redshift based on the cross-over in the equation of state for the 0.5 eV QCD axion dark matter, is a crucial test for the proposed cosmology. Together with the striking agreement between the predicted and observed values for the present energy density of cold dark matter in the universe, shown in Eqs.~(\ref{eq:axion_lambda}) and (\ref{eq:axion_lambda_planck}), this agreement on large-scale structure formation provides compelling evidence that the proposed 0.5 eV QCD axion dark matter cosmology is compatible with early-universe observations. In this way, we deepen the proposed resolution of the Hubble tension, sketched in Section \ref{sec:intro}, by ensuring that the agreement between the standard $\Lambda{\rm CDM}$ cosmology and early-universe observations is not spoiled by the switch to ``cooling'' dark matter, in the form of 0.5 eV QCD axions.  
	
	\section{Big Bang Nucleosynthesis}
	\label{sec:bbn}
	\begin{figure}
		\label{fig:bbn}
		\centering
		\includegraphics[width=1.0\linewidth]{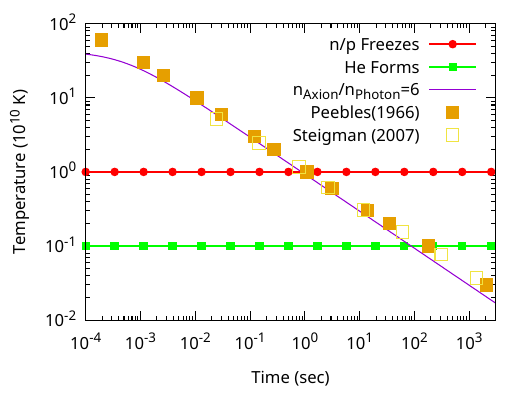}
		\caption{Cooling of the universe in the first three minutes after the Big Bang for a mix with six 0.5 eV QCD axions for each photon (violet line, \textcolor{violet}{\textemdash}) and for the standard mix of photons and light leptons (filled and unfilled boxes) both reach the freeze-out of the neutron-to-proton ratio (red horizontal line with dots) around the same time, but the standard mix \citep{peebles1966,steigman2007} takes roughly twice as long to start forming helium nuclei (green horizontal line with boxes). } 
	\end{figure}
	
	As the expanding universe cools in the first three minutes after the Big Bang (Fig.~3), the mix of particles with a light-like equation of state sets the time, $t_d$, that it takes to ``turn off'' the photodisintegration of the deuteron, $d$, so that the nuclei of light chemical elements beyond hydrogen, mainly helium, can form \citep[pgs.~184--186]{peebles1993}. Remarkably, the resulting ``primordial'' helium abundance (by mass), $Y_{\rm P}\approx0.3$, does not depend much on the baryon number density, $n_B(T_d)$, at the ``turn-off'' temperature, $T_d\approx10^{9}~{\rm K}$ \citep[pg.~129]{peebles2020}. Instead, the main constraints on baryon abundance from Big Bang nucleosynthesis (BBN) come from comparisons with observations of the ``primordial'' deuterium abundance (by number), $n_d/n_p=2.55~(3)\times10^{-5}$ \citep{cooke2018}, based on quasar absorption line spectroscopy. 
	
	Indeed, doubling the baryon number density, $n_B(T_d)$, compared to the standard cold dark matter cosmology value, as we propose to do in Eq.~(\ref{eq:ns-baryon}), ensures a nearly complete conversion of deuterons into helium, and, thus, a nearly complete {\em absence} of primordial deuterium \citep{cooke2018}, for the timing of the turn-off, $t_d(\rm{\Lambda CDM})\approx200~{\rm sec}$ \citep{peebles1966,steigman2007}, set by the standard mix of light leptons and photons. Yet, with the appropriate mix of six axions for every photon made in the Big Bang (See Section \ref{sec:hubble} and Appendix \ref{sec:appendix_A}), the turn-off time, $t_d\approx 100~{\rm sec}$, comes nearly twice as fast as it does with the standard mix (See Fig.~3). Thus, the value of the product that determines the deuterium abundance, $n_B(T_d)\langle\sigma v\rangle t_d \sim1$ \citep[pg.~141--142]{peebles1993}, lands in about the same place, preserving the agreement with observations, where $\langle\sigma v\rangle$ is the thermal average of the product of the neutron capture cross-section, $\sigma$, with the nucleon relative velocity, $v$.
	
	Still, the change in the time of the turn-off, $t_d$, has observable consequences due to the change in the gap in time, $t_d-t_n$, between the start of BBN, at $t_d$, and the ``freeze-out,'' at $t_n$, of the weak nuclear reactions between light leptons and nucleons (Fig.~3). During this gap, the decay of free neutrons with life-time, $\tau_n\approx878~{\rm sec}$ \citep{codata2025}, cuts down the neutron-to-proton ratio \citep[pg.~184--186]{peebles1993}. By cutting $t_d$ in half, while leaving $t_n$ more or less unchanged, we shorten the gap, $t_d-t_n$, by about ten percent of the free neutron lifetime, $\tau_n$, so that about ten per cent more neutrons make it to the start of BBN, at $t_d$, after which essentially all the surviving free neutrons end up as primordial helium \citep[pg.~128--129]{peebles2020}: We thus expect a roughly ten percent larger value for the primoridal helium abundance, $Y_{\rm P}\approx 0.27$, compared to the standard mix of light leptons and photons, $Y_{\rm P}({\rm \Lambda CDM})\approx0.25$ \citep{primat2018,parthenope2021}.  
	
	\section{Physical Picture}
	\label{sec:bkg}
	The proposed physical picture for dark matter, dark energy, and the Big Bang at the heart of 0.5 eV QCD axion cosmology begins with an observation by Zel'dovich in 1968 \cite{zeldovich1968}. Stimulated by some of the first quasar absorption line spectroscopy \citep{burbidge1967}, a number of astrophysicists suggested a value for the cosmological constant, $\Lambda\approx+5\times10^{-56}~{\rm cm}^{-2}$ \citep{kardashev1967,shklovskii1967,petrosian1967}. In this context, Zel'dovich observed that the gravitational binding energy, $U_{\Lambda}$, generated by a massless particle and antiparticle, each with the kinetic energy, $E$, separated by the distance, $R=\hbar c/E$,
	\begin{eqnarray}
		\label{eq:dark_energy}
		U_{\Lambda}&=&N_UG\frac{(E/c^2)^2}{R}\nonumber\\
		&=&N_U\frac{E^3}{E_{\rm Pl}^2}, 
	\end{eqnarray}
	with $N_U$ a dimensionless constant of order $\mathcal{O}(10^{-1})$, and having pair number density, 
	\begin{eqnarray}
		\label{eq:nlambda}
		n_{\Lambda}&=&N_n\frac{1}{R^3}\nonumber\\
		&=&N_n\left(\frac{E}{\hbar c}\right)^3,
	\end{eqnarray}
	with the dimensionless constant, $N_{n}$, of order $\mathcal{O}(10^{-1})$ to prevent Bose-Einstein condensation of the particle-antiparticle pairs, gives cosmological ``dark energy'' with energy density, $u_{\Lambda}=U_{\Lambda}n_{\Lambda}$, of the right size to explain the value of the cosmological constant, 
	$\Lambda=8\pi Gu_{\Lambda}/c^4$,
	\begin{eqnarray}
		\label{eq:dark_energy_density}
		u_{\Lambda}	&=& N_UN_n\left(\frac{E^3}{E_{\rm Pl}^2}\right)\left(\frac{E}{\hbar c}\right)^3\nonumber\\
		&=&1.874~708~(41)\times10^{-8}~{\rm erg~cm^{-3}}\nonumber\\
		&\times&\left(\frac{N_U}{0.1}\right)\left(\frac{N_n}{0.1}\right)\left(\frac{E}{105~{\rm MeV}}\right)^6,
	\end{eqnarray}	
	provided that the kinetic energy, $E$, is of the order of $\mathcal{O}(m_pc^2\times10^{-1})$, where $m_pc^2\approx 940~{\rm MeV}$ \cite{codata2025} is the rest-mass energy of the proton and $E_{\rm Pl}=\sqrt{\hbar c^5/G}=1.220~890~(14)\times10^{22}~{\rm MeV}$ \cite{codata2025} is the Planck energy scale.

	In this historical context, the key insight is simply to regard the particle-antiparticle pairs, which are responsible for the dark energy described by the cosmological constant, as forming the dark matter in the universe. The Big Bang, then, is just the moment when massless matter and antimatter bind to form the dark matter, in the form of these particle-antiparticle pairs. As the universe expands, the number density of dark matter, $n_A(t)\propto 1/a(t)^3$, dilutes with the volume of the expanding universe given by the scale factor of the universe, $a(t)$, at time $t$ after the Bang, and, thus, the kinetic energy of the massless matter and antimatter inside the dark matter, $E(t)\propto a(t)$, must grow to keep constant in time the value of the cosmological constant, $\Lambda\propto n_A(t)E(t)^3$.
	
Crucially, the dark energy described by the cosmological constant is not what binds the particle-antiparticle pair together to form dark matter. Instead, it is the dark matter rest-mass energy, $m_Ac^2$, which binds the particle-antiparticle pair. This rest-mass energy comes from the strong nuclear force described by quantum chromodynamics (QCD, see Appendix \ref{app:ma}). 
	
	Starting from the value, $\check{E}$, at the Big Bang, this kinetic energy reaches its peak value, $\hat{E}$, at future infinity when the dark matter dissolves and the universe begins contracting. For much of the past four decades, astrophysicists have looked to the Big Bang as a kind of cosmic accelerator \cite{euclid2018,cosmic_collider2017} for reaching collision energies sufficient to probe phenomena beyond the standard model of particle physics and the standard cold dark matter cosmology \citep[pg.~161]{lederman1989}. Yet, it is a simple historical fact that the last generation of the ``classic'' particle accelerators, known as synchrocyclotrons and built until the 1950s, already reached beam energies sufficient to make pairs of muons \citep[pgs.~93--94]{lederman1989}, and, thus these classic accelerators already achieved, seven decades ago, collision energies greater than those typically experienced in the Big Bang, where the typical kinetic energy of radiation, $\check{E}$, was on the order of, but smaller than, the rest-mass energy of the muon $m_{\mu}c^2\approx106~{\rm MeV}$.

	Following up on Zel'dovich's 1968 insight, we now estimate the typical kinetic energy per particle of radiation at the Big Bang, $\check{E}$, using a simple matching of the kinetic energy density, $\check{u}_R=(6+1)\check{u}_{\gamma}$, of the radiation formed in the Big Bang  --- six axions for every photon --- with the potential energy density released from the QCD vacuum, $\check{u}_{\rm QCD}$, by the ``jump'', $\Delta \theta_A=\pi$, in the axion angle, $\theta_A$, from the conformally invariant value, $\theta_A=\pi$, before the Bang, to the ``confining'' value, $\theta_A=0$, in our universe \citep{thooft1976a,thooft1976b,peccei1977a,peccei1977b}. Before the Bang, the critical soup of massless matter and antimatter interacts strongly and reaches thermal equilibrium. After the Bang, the radiation forms a Bose-Einstein gas with typical kinetic energy per particle, $\check{E}=x_k k\check{T}$, described by the Planck black-body law with the Wien displacement constant, $x_k=2.821~439~372$ \citep[pg.~159]{peebles1993}, at the temperature, $\check{T}$, determined by the matching:
	\begin{eqnarray}
		\label{eq:yT_derive}
		(6+1)\check{u}_{\gamma}=\left(\frac{\Delta\theta_A}{6}\right)^2\frac{\chi_{\rm QCD}}{(\hbar c)^3},
	\end{eqnarray}
	where $\check{u}_{\gamma}=(\pi^2/15)\left(kT_{\pi}\right)^4/(\hbar c)^3$ is the photon energy density \citep[pg.~137]{peebles1993}, and $\chi_{\rm QCD}$ is the topological susceptibility of the QCD vacuum \cite{thooft1976a,thooft1976b}.
	
	Using the best present estimate from chiral perturbation theory (CPT), $\sqrt{\chi_{\rm QCD}}=5690~(50)~{\rm MeV^2}$ \citep{gorghetto2019}, and the solution to the matching condition, $k\check{T}=y_T(\chi_{\rm QCD})^{1/4}$, with $y_T=(5/84)^{1/4}$, from Eq.~(\ref{eq:yT_derive}), we find the value of the kinetic energy at the Bang to be on the right order-of-magnitude to realize Zel'dovich's 1968 insight into the origin of the cosmological constant in terms of gravitational ``dark energy'' within particle-antiparticle pairs which we now interpret as 0.5 eV QCD axion dark matter:
	\begin{equation}
		\label{eq:echeck-cpt}
		\check{E}=(\chi_{\rm QCD})^{1/4}x_ky_T=105.12~(46)~{\rm MeV}.
	\end{equation} 
	 Indeed, this value for $\check{E}$ enters into Zel'dovich's estimate for the dark energy density, $u_{\Lambda}$, in Eq.~(\ref{eq:dark_energy_density}), together with the exact values for the dimensionless constants, $N_U=1/(2\pi^2)=0.050~660~592$, and, $N_n=6\times 2! \zeta(3)/(\pi^2x_k^3)=0.065~072~004$, derived in Appendix \ref{sec:appendix_A}, to give an estimate for the cosmological constant within 0.5 eV QCD axion cosmology:
	 \begin{eqnarray}
	 	\Omega_{\Lambda}h^2({\rm CPT})&=&\frac{8\pi G}{3c^4}\left(\frac{h}{H_0}\right)^2u_{\Lambda}\nonumber\\
	 	&=&0.368~8~(97),
	 \end{eqnarray} 
where the cosmological time-scale is, $h/H_0=3.085~677~581\times10^{17}~{\rm sec}$ \cite{cosmo-time}. Combining this estimate with the first-principles high-precision prediction for the value of the Hubble constant, $h=0.720~673~(79)$, given in Eq.~(\ref{eq:hubble}), we find rough agreement with the {\em Planck} estimate for the cosmological constant, $\Omega_{\Lambda}(Pl)=0.692~(5)$ \cite{planck2020,planck_baseline}:
\begin{equation}
	\label{eq:cpt-lambda}
	\Omega_{\Lambda}({\rm CPT})=0.710~(19).
\end{equation}

\section{Cosmological Parameters}
\label{sec:hubble}

Within the physical picture for the nature of dark matter, dark energy, and the Big Bang presented in section \ref{sec:bkg}, the conformal invariance of the critical liquid of massless matter and antimatter that emerges when dark matter dissolves and dark energy disappears at future infinity implies a scale-invariant form for the spectrum, $P(k)$, of primordial energy density fluctuations at the ``next'' big bang \citep{harrison1970a,zeldovich1972}. Technically, this means the spectrum follows a power-law, $P(k)=Ak^{n_s}$, for spatial frequency, $k$, that is much lower than the frequency-scale, $H_0(z_{\pi})/c$, set by the expansion rate of the universe, $H_0(z_{\pi})$, at the redshift of the Big Bang, $z_{\pi}\approx T_{\pi}/T_{\gamma}$, where $T_{\gamma}=2.7255~(6)~{\rm K}$ \cite{fixsen} is the temperature of the cosmic microwave background. Furthermore, the spectral index, $n_s$, is required by conformal symmetry to take the scale-invariant value \citep{harrison1970a,zeldovich1972}:
\begin{equation}
	n_s=1.
	\label{eq:n_s}
\end{equation} 

Following an insight into this conformal cross-over that was first formulated by Penrose \cite{penrose2016}, we apply a simple conformal rescaling of the Ricci curvature scalar to match the value at future infinity, $\hat{R}$, to that of the Big Bang, $\check{R}$ \citep[pg.~123]{penrose1986},
\begin{eqnarray}
	\label{eq:ricci-here}
	\hat{R}\hat{E}^2 =  \check{R}\check{E}^2,
\end{eqnarray}
where the kinetic energy per particle is $\hat{E}$ at future infinity and $\check{E}$ at the Big Bang. Further, we require that the conformal rescaling factors, $\hat{E}$ and $\check{E}$, satisfy the see-saw relation, or ``reciprocal hypothesis'' \cite{penrose2016}:
	\begin{equation}
		\label{eq:see-saw}
		\hat{E}\check{E}=f_A^2,
	\end{equation}
where $f_A$ is the decay constant of the 0.5 eV QCD axion dark matter \cite{wilczek1978,weinberg1978}. Finally, the physical picture for future infinity --- in terms of dark matter dissolving and dark energy disappearing --- suggests that we identify the axion decay constant, $f_A$, with the energy-scale set by the gravitational ``dark energy'' within the dark matter at future infinity \cite{zeldovich1972}:
\begin{equation}
	\label{eq:fA-dark-energy}
	f_A=\frac{G\left(\hat{E}/c^2\right)^2}{\hbar c/\hat{E}}=\frac{\hat{E}^3}{E_{\rm Pl}^2}.
\end{equation}   
	
Remarkably, the matching condition in Eq.~(\ref{eq:ricci-here}), the see-saw relation in Eq.~(\ref{eq:see-saw}), and the identification in Eq.~(\ref{eq:fA-dark-energy}), suffice to determine --- from first-principles and with high-precision --- the values for $f_A$, $\hat{R}$, and $\check{R}$, which in turn fix the values of the cosmological parameters for the energy density of cold dark matter, $\Omega_Ah^2$, the baryon abundance, $\Omega_Bh^2$, and the cosmological constant, $\Omega_{\Lambda}h^2$. The key point is that the field equations of general relativity with cosmological constant link the value of the Ricci curvature scalar at future infinity to the cosmological constant, $\hat{R}=4\Lambda$, and the value at the Big Bang to the trace of the baryon stress-energy tensor, $\check{R}=(8\pi G/c^4) \check{T}_B$ \citep[pg.~20]{penrose1986}. Using the dark energy density expression for the cosmological constant, given in Eq.~(\ref{eq:dark_energy_density}), it is then straightforward to derive the following functional identity satisfied by the temperature of the Big Bang, $T_{\pi}$:
\begin{equation}
	\label{eq:functional-equation}
	\frac{12}{\pi^2}x_k^{7/5}\left(\frac{kT_{\pi}}{E_{\rm Pl}}\right)^{2/5}=x_p e^{-x_p}J(x_p)+x_n e^{-x_n}J(x_n),
\end{equation}        
where the nucleon masses, $m_N$, set the ratios, $x_N=m_Nc^2/(kT_{\pi})$, for $N=p,~n$, which in turn fix the value of the integral, $J(x_N)=(1/2\zeta(3))\int_0^{\infty} dx e^{-x}\sqrt{x(x+2x_N)}$. 

We have solved the functional identity in Eq.~(\ref{eq:functional-equation}) with the following result:
\begin{equation}
	\label{eq:big-bang-temperature}
k\check{T} = 36.971~526~1~(84)~{\rm MeV}.
\end{equation}
Here the 0.23 part per million (ppm) uncertainty is set by the 11 ppm uncertainty in the Planck energy-scale, $E_{\rm Pl}=\sqrt{\hbar c^5/G}=1.220~890~(14)\times10^{22}~{\rm MeV}$, coming from the best present determination of the universal gravitational constant, $G=6.674~08~(15)\times10^{-8}~{\rm g^{-1}~cm^{3}~s^{-2}} $ \citep{codata2025}. Concretely, the sensitivity of the Big Bang temperature to the Planck energy-scale was found to be given by the following,
\begin{equation}
	\label{eq:temperature-planck}
s=\frac{E_{\rm Pl}}{kT_{\pi}}\frac{d(kT_{\pi})}{dE_{\rm Pl}} \approx -0.0205.
\end{equation}

We can then determine the kinetic energy per particle at the Big Bang to 0.23 ppm,
\begin{equation}
	\label{eq:e-check}
	\check{E}=x_kkT_{\pi}=104.312~919~(24)~{\rm MeV}.
\end{equation}	
A little more care is needed to determine the kinetic energy per particle at future infinity,
\begin{equation}
	\label{eq:e-hat}
	\hat{E}=\check{E}^{1/5}E_{\rm Pl}^{4/5}=1.183~066~(11)\times10^{18}~{\rm MeV},
\end{equation}	
where the fractional uncertainty is, $\Delta\hat{E}/\hat{E}=|(1/5)s+4/5|\Delta E_{\rm Pl}/E_{\rm Pl}=8.9~{\rm ppm},$ due to the sensitivity, $s$, given in Eq.~({\ref{eq:temperature-planck}). Similarly, the axion decay decay constant follows,
	\begin{equation}
		\label{eq:fA}
		f_A=\check{E}^{3/5}E_{\rm Pl}^{2/5}=1.110~896~2~(48)\times10^{10}~{\rm MeV},
	\end{equation}
with fractional uncertainty, $\Delta f_A/f_A=|(3/5)s+2/5|\Delta E_{\rm Pl}/E_{\rm Pl}=4.3~{\rm ppm}$.

The cosmological parameter, $\Omega_{\Lambda}h^2$, can then be predicted from first-principles with high-precision using the dark energy density expression in Eq.~(\ref{eq:dark_energy_density}),
\begin{eqnarray}
	\label{eq:cosmo-constant-here}
	\Omega_{\Lambda}h^2&=&\frac{16\zeta(3)x_k^3}{\pi^3}\frac{(kT_{\pi})^6}{E_{\rm Pl}^4}\left(\frac{h}{H_0}\frac{1}{\hbar}\right)^2\nonumber\\
	&=&0.352~174~(16),
\end{eqnarray}
with fractional uncertainty, $\Delta\Omega_{\Lambda}h^2/(\Omega_{\Lambda}h^2)=|6s-4|\Delta E_{\rm Pl}/E_{\rm Pl}=46~{\rm ppm}$. To determine the baryon abundance, $\Omega_Bh^2$, we first find the baryon-to-photon number density ratio, $\eta$,
\begin{equation}
	\label{eq:eta}
\eta=e^{-x_p}I(x_p)+e^{-x_n}I(x_n)=1.343~889~8~(64)\times10^{-9},
\end{equation}
where the integral is $I(x_N)=(1/2\zeta(3))\int_0^{\infty}dx e^{-x}(x+x_N)\sqrt{x(x+x_N)},$ and the fractional uncertainty was $\Delta\eta/\eta=(\eta/E_{\rm Pl})(d\eta/dE_{\rm Pl})(\Delta E_{\rm Pl}/E_{\rm Pl})=4.8~{\rm ppm}.$ Similarly, to fix the present energy density of cold dark matter, $\Omega_Ah^2$, we first find the axion rest-mass energy, $m_Ac^2=\sqrt{\chi_{\rm QCD}}/f_A$ \cite{wilczek1978,weinberg1978}, using the expression for the axion decay constant, $f_A$, in Eq.~(\ref{eq:fA}), and the relation, $\sqrt{\chi_{\rm QCD}}=y_T^{-2}(kT_{\pi})^2=5602.597~2~(25)~{\rm MeV}^2$, from Eq.~(\ref{eq:yT_derive}),
	\begin{equation}
		\label{eq:mA}
		m_Ac^2=x_k^{-3/5}y_T^{-2}(kT_{\pi})^{7/5}E_{\rm Pl}^{-2/5}=0.504~331~0~(24)~{\rm eV},
	\end{equation}
with fractional uncertainty, $\Delta m_Ac^2/(m_Ac^2)=|(7/5)s-2/5|\Delta E_{\rm Pl}/E_{\rm Pl}=4.8~{\rm ppm}.$

Finally, we evaluate the cosmological parameter, $\Omega_Bh^2$, within 0.5 eV QCD axion cosmology,
\begin{equation}
	\label{eq:baryon-abundance}
	\Omega_Bh^2=\frac{8\pi Gm_p}{3}\left(\frac{h}{H_0}\right)^2n_{\gamma}\eta(1+\epsilon)=0.049~243~(32),
\end{equation}
with the 0.66 per mille precision set by the cosmic microwave background photon number density, $n_{\gamma}=2\zeta(3)(kT_{\gamma}/(\hbar c))^3=410.73~(27)~{\rm cm^{-3}}$ \citep[pg.~134-137,158--159]{peebles1993}. Here, the Big Bang nucleosynthesis correction, $\epsilon=(4m_{\alpha}/m_p-1)Y_P=1.876~(34)\times10^{-3}$, is estimated using the value of the primordial helium abundance, $Y_P\approx Y_{\rm HB}=0.272~(5)$, suggested by observations of horizontal branch stars (See Eq.~(\ref{eq:yhb})), and the helion-to-proton mass ratio, $m_{\alpha}/m_p=3.972~599~690~252~(70)$ \cite{codata2025}. Similarly, the cosmological parameter, $\Omega_Ah^2$, can be predicted from first-principles to 0.66 per mille precision,
\begin{equation}
	\label{eq:axion-abundance}
	\Omega_Ah^2=\frac{8\pi G}{3c^2}\left(\frac{h}{H_0}\right)^2 6n_{\gamma}m_Ac^2=0.117~952~(78),
\end{equation}
where the fact that there are six axions for every photon in the CMB was used (See Appendix \ref{sec:appendix_A}). For a spatially flat, isotropic, and homogeneous expanding-universe solution to the field equations of general relativity with cosmological constant, the present expansion rate of the universe, $H_0=100h~{\rm km s^{-1} Mpc^{-1}}$, given in Eq.~(\ref{eq:hubble}), is determined by the cosmological parameters, $\Omega_Ah^2, \Omega_B h^2,$ and $\Omega_{\Lambda} h^2$ \cite[pgs.~100]{peebles1993},
	\begin{equation}
		h=\sqrt{\Omega_A h^2+\Omega_B h^2+\Omega_{\Lambda} h^2}=0.720~673~(79),
	\end{equation}
with the fractional uncertainty, $\Delta h/h=(\Omega_M/2)\Delta\Omega_Ah^2/(\Omega_Ah^2)=0.11~{\rm per~mille}$, set by the matter abundance, $\Omega_M=1-\Omega_{\Lambda}=0.321~92~(15)$.

	\section{Discussion}
	\label{sec:conc}
	We have presented a physical picture for the nature of dark matter, dark energy, and the Big Bang. Dark matter is made of quantum chromodynamic (QCD) axions with rest-mass energy around 0.5 eV. Dark energy is the gravitational attraction between the particle of matter and the particle of antimatter, both moving with the speed of light, that bind to make the dark matter in the moment of the Big Bang. 
	
	Before the Bang, massless quarks and leptons, together with their antiparticles, move in a conformally invariant critical soup, with neither dark matter nor dark energy. Correspondingly, at future infinity, the axion dark matter dissolves and the dark energy disappears. Until the next big bang, the universe shrinks and the temperature rises.
	
	In the present universe, the rate with which space grows over time is predicted within the resulting 0.5 eV QCD axion cosmology with high-precision from first-principles. Remarkably, the predicted value does not disagree sharply with the best present determination, by the SH0ES collaboration in 2025, using late-universe observations of relatively nearby Type Ia supernovae with low redshift. In this sense, the proposed cosmology has the potential to resolve the tension between these late-universe observations and the standard cold dark matter cosmology with cosmological constant (${\rm \Lambda CDM}$).
	
	However, to realize such a resolution, the proposed cosmology must also pass a battery of observational tests which draw on a wide range of astrophysical phenomena, including those at high redshift. Remarkably, the predicted values for the cosmological constant and cold dark matter abundance --- from first-principles with high-precision within 0.5 eV QCD axion cosmology --- do not sharply disagree with the best present determination, by the {\em Planck} collaboration in 2018, using early-universe observations of cosmic microwave background (CMB) fluctuations, baryon-acoustic oscillations (BAO), and cosmic shear. Similarly, the timing of the onset of large-scale structure formation within the proposed cosmology also seems to roughly agree with the {\em Planck} determination, while the change in the timing of the onset of Big Bang nucleosynthesis --- roughly twice as fast as ${\rm \Lambda CDM}$ --- combines with the roughly twice as large value for the baryon abundance to yield about the same primordial deuterium abundance, in agreement with observations using quasar absorption line spectroscopy from near cosmic noon. 
	
	Moving forward, a straightforward calculation of Big Bang nucleosynthesis within 0.5 eV QCD axion cosmology promises to yield high-precision predictions from first-principles for the primordial deuterium and helium abundances. Similarly, a straightforward calculation of large-scale structure formation within 0.5 eV QCD cosmology offers the prospect of estimating the spectrum of energy density fluctuations in the early-universe from first-principles with high-precision. In this way, the rough agreement between 0.5 eV QCD axion cosmology and early-universe observations, presented here, can be tested directly.
	
	In passing we note the rough agreement between the prediction from first-principles to high-precision within 0.5 eV QCD axion cosmology for the value of the topological susceptibility of the QCD vacuum, $\chi_{\rm QCD}$, above Eq.~(\ref{eq:mA}), and the value inferred with chiral perturbation theory (CPT) \cite{gorghetto2019},
	\begin{eqnarray}
		\label{eq:chi-qcd-compare}
		\sqrt{\chi_{\rm QCD}}&=& 5602.597~2~(25)~{\rm MeV^2}\nonumber\\
		\sqrt{\chi_{\rm QCD}}({\rm CPT})&=& 5690~(50)~{\rm MeV}^2.
	\end{eqnarray}
	It is worth emphasizing that the precision of the prediction for the axion rest-mass energy, $m_Ac^2\approx0.504~331~0~(24)~{\rm eV}$, given in Eq.~(\ref{eq:mA}), is set by the best present determination of the universal gravitational constant, $G$. Therefore, within the context of 0.5 eV QCD axion cosmology, a measurement of $m_Ac^2$ to better than the 4.8 ppm precision of the current prediction would, in effect, constitute an improved determination of $G$.
	
	Finally, we discuss the attempts to ``constrain'' or ``exclude'' 0.5 eV QCD axion dark matter using astrophysical phenomena and the coupling of the axion to photons, electrons, and nucleons. For axion-photon coupling, the most recent attempts are based on the spontaneous two-photon decay of 0.5 eV QCD axions in the dark matter ``halo'' of the Milky Way \citep{pinetti2025,pinetti2025_prl,roy2025}, while a more established approach looks for axion production from gamma rays in the cores of horizontal branch stars within globular clusters \citep{ayala2014}. Efforts involving axion-electron coupling go back five decades \citep{sato1975,sato1978}, focused mainly on red giants and white dwarfs, while attempts invoking axion-nucleon coupling have instead looked at pulsars and supernova 1987A (See ref.~\cite{raffelt2024} for a recent review).
	
	\begin{figure}
		\centering
		\includegraphics[width=1.0\linewidth]{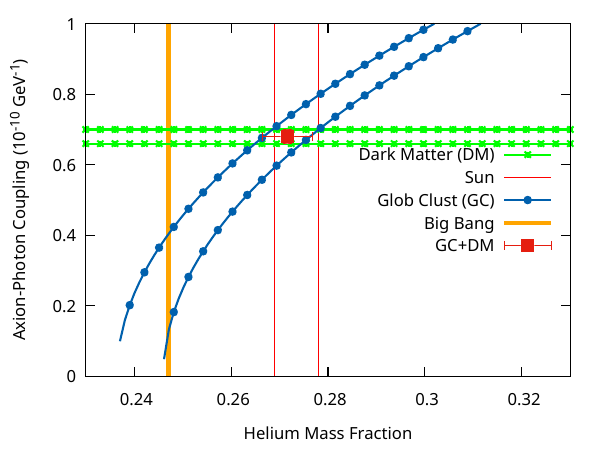}
		\label{fig:helium}
		\caption{Axion-photon coupling grows with initial core helium mass fraction in simulations of stellar evolution for horizontal branch stars constrained by observations of globular clusters (curved blue lines with dots) \citep{ayala2014}. The predicted value for the axion-photon coupling strength (horizontal green lines with boxes), given in Eq.~(\ref{eq:ga}), combines with this empirical-computational relation to give a prediction for the helium mass fraction (red box with whiskers) that differs by $5.0\sigma$ from the primordial helium abundance expected within the standard cold dark matter cosmology with cosmological constant ($\Lambda {\rm CDM}$) based on simulations of Big Bang nucleosynthesis (thick vertical orange line) \citep{parthenope2021,primat2018}, but that agrees rather well with the best present estimates for the initial core helium mass fraction in the Sun (pair of thin vertical red lines) \citep{piersanti2007,serenelli2010}.} 
	\end{figure}
	
	Figure 4 shows the empirical-computational relation between the predicted axion-photon coupling strength (See Appendix \ref{sec:axion-photon}), 	
	\begin{equation}
		\label{eq:ga}
		g_{A\gamma\gamma}=g_{10}\times 10^{-10}~{\rm GeV}^{-1}=0.68~(2)\times10^{-10}~{\rm GeV^{-1}},
	\end{equation}
	 and the initial core helium mass fraction, $Y_{\rm HB}$, for stars that are now on the horizontal branch (HB) of the color-magnitude diagram for globular clusters \citep{ayala2014}:
	\begin{eqnarray}
		\label{eq:glob_clust}
		R=6.26Y_{\rm HB}-0.41g_{10}^2-0.12,
	\end{eqnarray}
	where $R=N_{\rm HB}/N_{\rm RGB}=1.39~(3)$ was the observed ratio of the number of HB stars, $N_{\rm HB}$, to the number of stars on the red giant branch (RGB), $N_{\rm RGB}$, of the color-magnitude diagram in 39 globular clusters. Combining the empirical-computational relation in Eq.~(\ref{eq:glob_clust}) with the predicted value for the axion-photon coupling in Eq.~(\ref{eq:ga}), we find a percent-level precise estimate for the initial core helium mass fraction in these HB stars that formed in the first billion years after the Big Bang:
	\begin{eqnarray}
		\label{eq:yhb}
		Y_{\rm HB}=0.272~(5),
	\end{eqnarray} 	
	which differs by $5\sigma$ with the best present estimate of the ``primordial'' helium abundance, $Y_P({\rm \Lambda CDM})=0.247~09~(17)$ \citep{primat2018,parthenope2021}, within the standard cold dark matter cosmology, but that roughly matches the best present estimates for the initial core helium mass fraction in the Sun, $Y_{\odot}=0.273~(6)$ \citep{serenelli2010,piersanti2007}. As discussed in Section \ref{sec:bbn}, the mix of axions and photons that dominates the expansion of the universe leading up to Big Bang nucleosynthesis, within the proposed physical picture, yields about ten percent more primordial helium, $Y_P\approx0.27$, than the standard mix of light leptons and photons, and, in this way, resolves the $5\sigma$ tension between $Y_{\rm HB}$ in Eq.~(\ref{eq:yhb}) and $Y_P({\rm \Lambda CDM})$.  
	
	Similarly, the recent attempts to ``rule out'' axion dark matter with the rest-mass energy in Eq.~(\ref{eq:mA}) and the axion-photon coupling in Eq.~(\ref{eq:ga}) using ``blank-sky'' observations of the dark matter ``halo'' of the Milky Way at the near-infrared wavelength $2\lambda_A=2h/m_Ac\approx4.9~{\rm \mu m}$, corresponding to the decay of the axion into a pair of photons, subtract a ``smooth astrophysical background'' by fitting a quartic polynomial to the near-infrared spectrum within a window of width $\Delta \lambda\approx0.2~{\rm \mu m}$ \citep{pinetti2025}. However, the proposed physical picture implies that 0.5 eV QCD axion dark matter is in thermal equilibrium with the cosmic microwave background (See Section \ref{sec:hubble} and Appendix \ref{sec:appendix_A}), and, thus, the axions in the dark matter ``halo'' of the Milky Way have velocity dispersion, $\sigma_v$, fixed by the CMB temperature, $T_{\gamma}$, and by the 0.5 eV QCD axion dark matter rest-mass energy, $m_Ac^2$, leading to an expected line-width, $2\Delta\lambda_A$, that is roughly the same size as the window used to subtract the ``smooth astrophysical background'':
	\begin{eqnarray}
		\label{eq:vel_disp}
		2\Delta\lambda_A&=&2\lambda_A\frac{\sigma_v}{c}=2\lambda_A\sqrt{\frac{3kT_{\gamma}}{m_Ac^2}}\approx0.2~{\rm \mu m}.
	\end{eqnarray}
	Instead of ``constraining'' the 0.5 eV QCD axion, the analysis by ref.~\cite{pinetti2025} in effect rules out the standard cold dark matter velocity dispersion, $\sigma_v({\rm \Lambda CDM})\approx 200~{\rm km~sec^{-1}}$ \citep{goodman1985}, which is based on the assumption of gravitational equilibrium with ``bright'' matter, such as OB associations, H II regions, and 21 cm clouds, and is much smaller than the velocity dispersion, $\sigma_v\approx11,000~{\rm km~sec^{-1}}$, given in Eq.~(\ref{eq:vel_disp}) by thermal equilibrium with the CMB, within the physical picture for dark matter proposed here. 
	
	Turning to axion-electron coupling, the key point is that there is no ``tree-level'' or ``bare'' coupling between the axion and electron \citep{peccei1989}, just as there is no such coupling between the neutral pion, $\pi^0$, and the electron \citep{husek2024}. Instead, the ``leading order'' axion-electron coupling appears only at ``loop-level,'' and is suppressed by the small factor, $(\frac{\alpha}{2\pi})^2\sim10^{-6}$ \citep{codata2025}, compared to the ``bare'' value used to derive astrophysical ``constraints'' and ``exclusions'' \citep{sato1975,sato1978,raffelt2024}. The upshot is that the effect of axions on the astrophysical phenomena, such as white dwarf cooling and red giant luminosity, is smaller than the astrophysical uncertainties entering the analysis, and, thus, no significant ``constraints'' or ``exclusions'' emerge from the comparison with astrophysical observations. 
	
	Finally, regarding the axion-nucleon coupling, a dense bunch of nucleons makes a lot of 0.5 eV QCD axions due to this coupling, but, by the same token, such a dense nuclear soup is ``axionically thick,'' in the sense that a 0.5 eV QCD axion made by one nucleon is unlikely to ``escape'' the concentration without coupling to another nearby nucleon, as pointed out by Turner in 1988 \cite{turner1988}, in the analysis of the effect of axions on supernova 1987A. Remarkably, there is a relatively narrow window of QCD axion rest-mass energy, $0.05~{\rm eV}\le m_Ac^2\le 5~{\rm eV}$ \citep[pg.~42]{wenyin2023}, that brackets the predicted value in Eq.~(\ref{eq:mA}), in which the effects of the QCD axions completely dominate the flow of energy within the dense nuclear soup, in sharp contrast with the traditional assumption that axions play essentially no role in this flow, and this contrast is often expressed as an ``exclusion'' of the 0.5 eV QCD axion. However, given the apparent success of 0.5 eV QCD axion dark matter in resolving the Hubble tension, which is the main focus here, one is instead led, on the basis of this analysis of the axion-nucleon coupling, to question the traditional assumptions regarding the flow of energy within compact astrophysical bodies, such as pulsars, as well as the flow within the events that are thought to form them, such as Type II supernovae.     
	
	\begin{acknowledgments}
		Nadir Jeevanjee, Peter John Moraga, and Henry W. Lin asked probing questions, listened patiently, and offered helpful suggestions. The organizers of the 20th Patras Workshop on Axions, WIMPs, and WISPs generously included a talk by the author on an earlier version of this work and the discussions with the participants, particularly Javier Redondo, are gratefully acknowledged. Daria Mazura provided critical and constant support without which the work reported here would not have been possible. 	
	\end{acknowledgments}
	\appendix
	\section{Dark Energy and Number Density }
	\label{sec:appendix_A}
	Within the physical picture presented in Section \ref{sec:bkg}, the gravitational dark energy of 0.5 eV QCD axion dark matter, $U_{\Lambda}=N_UG(E/c^2)^2/R$, given in Eq.~(\ref{eq:dark_energy}), is generated by the kinetic energy, $E$, of a massless particle and anti-particle, each with de Broglie wavelength, $\lambda=hc/E=2\pi R$. The value of the dimensionless constant, $N_U=1/(2\pi^2)$, then follows from two assumptions. First, the axion is ``spin-less,'' in the sense that the plane of motion of the particle and anti-particle has equal probability amplitude to be oriented in any direction in three-dimensional space, and, second, the three-dimensional vector, ${\bf r}$, separating the particle and anti-particle, is related to twice the de Broglie wavelength, $D=2\lambda$, by the familiar geometric relation, $1/D=\langle 1/|{\bf r}|\rangle_{\Omega}$, for the angle-averaged plane-projected (``line-of-sight'') distance \citep[pgs.~338--339]{barger1995}:
	\begin{eqnarray}
		\label{eq:proj_dist}
		\frac{1}{D}&=&\left\langle \frac{1}{|\bf{r}|}\right\rangle_{\Omega}\nonumber\\
		&=&\frac{1}{4\pi}\int\frac{d\Omega}{r\sin \theta}\nonumber\\
		&=&\frac{1}{4\pi r} \int_0^{\pi} d\theta\int_0^{2\pi} d\phi\nonumber\\
		&=&\frac{\pi}{2r},
	\end{eqnarray}
	where, $r$ is the three-dimensional separation, $\Omega$ is the solid angle on the two-dimensional sphere of orientations of the orbital plane of the particle-antiparticle pair relative to some fixed ``line-of-sight,'' $\theta$ is the polar angle (co-altitude) of the line-of-sight to the normal vector of the plane, and $\phi$ is the azimuthal angle of this normal vector with respect to the line-of-sight.
	
	Using the geometric identity in Eq.~(\ref{eq:proj_dist}), it is then straightforward to derive the gravitational self-energy of the 0.5 eV QCD axion dark matter:
	\begin{eqnarray}
		\label{eq:nu_derive}
		U_{\Lambda}&=&G\frac{(E/c^2)^2}{r}\nonumber\\
		&=&\frac{2}{\pi}G\frac{(E/c^2)^2}{D}\nonumber\\
		&=&\frac{1}{2\pi^2}G\frac{(E/c^2)^2}{R}.
	\end{eqnarray}
	Comparing the last line of Eq.~(\ref{eq:nu_derive}) with the expression for the gravitational self-energy in Eq.~(\ref{eq:dark_energy}), we find the value of the dimensionless constant, $N_U=1/(2\pi^2)$. We emphasize that this result requires that the 0.5 eV QCD axion dark matter be a {\em spin-less} bound-state of particle and anti-particle, and, thus, the axion must obey Bose-Einstein statistics \citep[pg.~205]{landau_qm}, just as photons do \citep[pg.~134]{peebles1993}.  
	
	At the Big Bang, the average kinetic energy, $\check{E}=x_kkT_{\pi}\approx104~{\rm MeV}$, given in Eq.~(\ref{eq:e-check}), is much larger than the 0.5 eV QCD axion dark matter rest-mass energy, $m_Ac^2=0.5~{\rm eV}=5\times10^{-7}~{\rm MeV}$, given in Eq.~(\ref{eq:mA}). The axion number density at the Big Bang, $\check{n}_A$, is then determined by the Planck distribution function, just as the photon number density at the Big Bang, $\check{n}_{\gamma}$, is so determined \citep[pg.~134--137, 158--159]{peebles1993}, with the result
	\begin{eqnarray}
		\label{eq:nA_derive}
		\check{n}_A&=& \frac{g_A}{2} \frac{2!\zeta(3)}{\pi^2}\left(\frac{kT_{\pi}}{\hbar c}\right)^3\nonumber\\
		&=&6\check{n}_{\gamma}\nonumber\\
		&=&\left(\frac{T_{\pi}}{T_{\gamma}}\right)^3n_A,
	\end{eqnarray}
	where $n_A$ is the axion number density in the present universe, $T_{\gamma}$ is the temperature of the cosmic microwave background, and the statistical degeneracy factor, $g_A=12$, counts the twelve ``flavors'' of quarks and leptons in the standard model of particle physics \citep[pg.~128]{lederman1989}. Each flavor, $M$, of quark and lepton binds with an anti-particle of the same flavor and helicity to make the axion, $A_M$, with this flavor and with no spin (See Appendix \ref{sec:axion-photon} for further discussion on the structure of the axion), such that the axion number density at the Big Bang is given by the expression in Eq.~(\ref{eq:nlambda}), with the value of the dimensionless constant, $N_n=6\times2!\zeta(3)/(\pi^2x_k^3)$, fixed by the result given above in Eq.~(\ref{eq:nA_derive}). 
	
	\section{Axion-Photon Coupling Strength}
	\label{sec:axion-photon}
	The strength of the axion-photon coupling, $g_{A\gamma\gamma}$, that enters the Lagrangian density, $\mathcal{L}_{A\gamma}=g_{A\gamma\gamma}\phi_A{\bf E}\cdot{\bf B}$, for the interaction between the axion field, $\phi_A$, and the dot product, ${\bf E}\cdot{\bf B}$, of the electric field, ${\bf E}$, and the magnetic field, ${\bf B}$, is determined by the decay constant of the axion, $f_A$, and by the dimensionless constant, $C_{A\gamma}=\mathcal{O}(1)$, whose value is of order one \citep{wilczek1978,weinberg1978}
	\begin{eqnarray}
		\label{eq:ga_basic}
		g_{A\gamma\gamma}&=&C_{A\gamma}\frac{\alpha}{2\pi}\frac{1}{f_A}\nonumber\\
		&=&1.045~471~0~(45)\times10^{-13}~{\rm MeV}^{-1}~C_{A\gamma}, 
	\end{eqnarray}   
	where $\alpha=e^2/(\hbar c)=1/(137.035~999~18~(2))$ \citep{codata2025} is the best present estimate for the dimensionless electromagnetic ``fine-structure'' coupling constant. To estimate the value of $C_{A\gamma}$, we use its relation with the light-quark mass ratio, $m_u/m_d=0.485~(19)$ \citep{fodor2016}, and with the electromagnetic-to-strong axial anomaly ratio, $E/N=8/3$ \citep{peccei1989,gorghetto2019},
	\begin{eqnarray}
		\label{eq:ca}
		C_{A\gamma}&=&\frac{E}{N} - \frac{2}{3}\frac{4m_d+m_u}{m_d+m_u}\nonumber\\
		&=&	0.65~(2),
	\end{eqnarray} 
	where the value of the anomaly ratio is set by the fact that all twelve flavors of ``normal'' quarks and leptons within the standard model of particle physics transform with the same axial charge, $Q_A=+1$, under the continuous axial $U(1)_A$ phase symmetry, $U_A(\alpha)=\exp(iQ_A\alpha)$, while the opposite axial charge, $Q_A=-1$, holds for ``mirror'' quarks and leptons which have the same correlation between chirality and helicity as ``normal'' antimatter (Right-handed mirror matter feels the weak nuclear force but left-handed mirror matter does not feel it; See ref.~\cite{lederman1989}, pg.~121--123): The 0.5 eV QCD axion dark matter, $A_M$, with a given flavor, $M$, is the unique quantum state formed by superposing the bound state of mirror matter and normal antimatter with the bound state of normal matter and mirror antimatter, such that the particle and the antiparticle with flavor $M$ inside the bound states have the same helicity and chirality, and such that the resulting quantum state for the axion, and hence the axion field, $\phi_A$, has the same discrete space-time symmetry properties as the dot product, ${\bf E}\cdot{\bf B}$, ensuring that the Lagrangian density, $\mathcal{L}_{A\gamma}$, is invariant under these transformations (charge-conjugation, space-inversion, and time-reversal; See ref.~\cite{lederman1989}, pg.~121--123). Finally, we combine Eq.~(\ref{eq:ga_basic}) and Eq.~(\ref{eq:ca}) to find the value for the axion-photon coupling strength given in Eq.~(\ref{eq:ga}).  	
	
\section{Axion Rest-Mass Energy}
\label{app:ma}	
The QCD axion dark matter rest-mass energy, $m_Ac^2\approx5.04\times10^{-7}~{\rm MeV}$, given in Eq.~(\ref{eq:mA}), comes fundamentally from the strong nuclear force described by quantum chromodynamics (QCD) \cite{thooft1976a,thooft1976b}:
\begin{equation}
	\label{eq:ma-qcd}
	m_Ac^2=2\hat{E}e^{-2\pi/\alpha_s(v)},
\end{equation}
where $\alpha_s(v)$ is the strong coupling constant evaluated at the weak nuclear force energy scale, $v=1/(\sqrt{2}G_F)^{1/2}\approx 2.46\times10^5~{\rm MeV}$ \cite{codata2025}, and $\hat{E}\approx1.18\times10^{18}~{\rm MeV}$, given in Eq.~(\ref{eq:e-hat}), is the kinetic energy at future infinity. The best present estimate of the strong nuclear force coupling constant, $\alpha_s(M_Z)$, evaluated at the neutral weak $Z$ boson rest-mass energy, $M_Z$, comes from lattice QCD \cite{hpqcd2010} (see also \cite{alpha2026}):
\begin{equation}
	\label{eq:alphas-qcd}
	\alpha_s(M_Z,{\rm lattice})=0.1183~(7)~[5900~{\rm ppm}].
\end{equation}
Using the value at the weak energy scale, $v$, for the strong coupling constant, $\alpha_s(v)=2\pi/\ln(2\hat{E}/m_Ac^2)=0.110~604~230~(27)~[{\rm 0.24~ppm}]$, determined by Eq.~(\ref{eq:ma-qcd} with the precision, $\Delta\alpha_s(v)/\alpha_s(v)=(6/5)(1-s)\Delta E_{\rm Pl}/E_{\rm Pl}=0.24~{\rm ppm}$ (See Section \ref{sec:hubble}), we find a high-precision first-principles prediction for its value, $\alpha_s(M_Z)$, at the rest-mass energy of the $Z$ boson, $M_Z=9.118~72~(21)\times10^{4}~{\rm MeV}$ \cite{codata2025}, using the renormalization group ``flow'' equation for QCD \cite{politzer1973,wilczek1973} with $n_f=12$ massless quark ``flavors'', corresponding to the six ``normal'' quark flavors within the standard model of particle physics \cite{codata2025} along with the six ``mirror'' quark flavors which appear within the axion (See Appendix \ref{sec:axion-photon}),
\begin{eqnarray}
	\label{eq:alphas-ours}
	\alpha_s(M_Z)&=&\frac{\alpha_s(v)}{1-\frac{\alpha_s(v)}{2\pi}\left( \left(\frac{11}{3}\times3\right)-\left(\frac{2}{3}\times n_f\right)\right) \ln\left(\frac{v}{M_Z}\right)}\nonumber\\
	&=&0.116~727~27~(15)~[{\rm 1.3~ppm}],
\end{eqnarray} 
which roughly agrees with the lattice QCD estimate in Eq.~(\ref{eq:alphas-qcd}), to within the percent-level uncertainty of the comparison, while the 0.5 eV QCD axion cosmology estimate is nearly $4600\times$ more precise.

\providecommand{\href}[2]{#2}\begingroup\raggedright
\endgroup

	\end{document}